\newif\iffiginclude
\figincludetrue          
\iffiginclude
\documentstyle[12pt,psfig]{article}
\else
\documentstyle[12pt]{article}
\fi

\font\tenrm=cmr10
\font\tenit=cmti10
\font\elevenbf=cmbx10 scaled\magstep 1
 1
 1

\begin{document}

\rightline {\bf UPR-596-T}
\rightline {\bf December 1993}

\begin{center}

\vglue 1.0cm

{ {\elevenbf        \vglue 10pt
  Supergravity Walls as Windows into New  Types of Vacua
\footnote{Contribution to
the Proceedings of the International Europhysics
Conference on High Energy Physics,
Marseille, July 21-28, 1993.}
 }
\vglue 1.0cm
{\tenrm MIRJAM CVETI\v C \\}
\baselineskip=13pt
{\tenit Department of Physics, University of Pennsylvania \\}
\baselineskip=12pt
{\tenit Philadelphia, PA 19104-6396, USA \\}}\vglue 1.0cm

\def\ge{\lower3pt\hbox{$\buildrel>\over -$}}
\def\le{\lower3pt\hbox{$\buildrel<\over-$}}%
\def\overtext#1{$\overline{#1}$}
\def\etal{{\it et al.}}
\def\to{\rightarrow}

\vglue 0.8cm
{\tenrm ABSTRACT}

\end{center}

\vglue 0.3cm
{\rightskip=3pc
 \leftskip=3pc
 \tenrm\baselineskip=12pt
 \noindent
Within four dimensional ($4d$)  $N=1$  supergravity theories
we present   extreme dilatonic domain wall solutions  with a  general
 overall coupling  $\alpha $ in the dilaton K\" ahler potential.
We concentrate on  extreme Type  I  walls,  which are static, planar
configurations, interpolating
between Minkowski space-time and  a new type of space-time with a
varying dilaton field.   $\alpha=0$  case yields  extreme ``ordinary''
supergravity walls  between Minkowski and anti-de Sitter space-times.
For $\alpha=1$  the walls are extreme ``stringy dilatonic'' walls of the
three level superstring vacua  interpolating between the constant
and  linear dilaton vacuum. Extreme  stringy dilatonic walls  ($\alpha =1$)
serve
as  a dividing line between the
walls ($0<\alpha<1$) with the  (planar) singularity covered
 by the horizon  and those ($\alpha>1$) with the   naked singularity.
Striking similarities  between the global space-time structure  of  such
 domain walls and and the one of the corresponding extreme  charged
black holes with  a  general dilaton coupling   are pointed out.}

\newpage
Over the last few years    vacuum domain walls
  in four-dimensional ($4d$)
$N=1$ supergravity theories have been addressed. The work stems from an earlier
discovery  of extreme ``ordinary''    walls \cite{CGR},\cite{CG}
and  a subsequent study  of
their global space-time structure \cite{CDGS},\cite{GIBBIII} . Such walls
are static, planar  walls, in general interpolating between  vacua of
non-equal, non-positive cosmological  constants.  Extreme Type I walls  with
the energy density $\sigma=\sigma_{ext}$
interpolate between  Minkowski and anti-de Sitter  vacua.
Furthermore,  extreme ordinary  domain
walls ($\sigma=\sigma_{ext}$) serve as a dividing line
between the non-extreme ($\sigma>\sigma_{ext}$, expanding bubbles
 with two insides) and
ultra-extreme  ($\sigma<\sigma_{ext}$, false vacuum decay bubbles) wall
solutions \cite{CGS},\cite{CGSII} .
Ordinary walls  provide a fertile ground to study
globally non-trivial space-times, which turn out to be
 closely related \cite{CDGS},\cite{CGS}\ to the ones of charged
(Reissner-Nordstr\"om) black holes, however,
without singularities.

On the other hand, ``stringy  dilatonic''   walls \cite{C}  correspond to
solutions specific to  $4d$ tree level superstring vacua. Here  along with the
the matter field, forming the wall, not only the metric, but
also the dilaton field  changes its value in the domain  wall background.
Extreme Type I   stringy dilatonic walls are  static, planar configurations
interpolating between the constant dilaton vacuum (Minkowski space-time)
and the linear dilaton vacuum. In the sigma model frame of the string
theory (string frame), the metric is flat
everywhere in the domain wall background.  In addition, the  global space-time
is
closely related to the one of the corresponding charged dilatonic black
holes \cite{GA},\cite{GM},\cite{GHS} .\footnote{For a review see Ref. \cite{H}
.}
Further work  which would shed light on non-extreme stringy
dilatonic wall solutions and their global space-time  is  needed.

Recently, it was pointed out \cite{CY} that in the case of more than one
dilaton  field, each of them respecting the non-compact $SU(1,1)$ symmetry in
its
subsector, the  extreme Type I domain wall
solutions  interpolate between the Minkowski space-time
and a new type of supersymmetric vacua which exhibit the naked  (planar)
 singularity.
This is the first example of classical supersymmetric ``solitons'' whose
induced
space-time exhibits naked singularities.

Here I would  like to present   extreme
dilatonic domain walls  with a {\it general} dilaton coupling within $4d$ $N=1$
 supergravity theory.   Dilaton field has its origin  as a real scalar
component of the
linear supermultiplet. Within the K\" ahler
superspace formalism  the dilaton field
$\phi$ is represented as a real part of the scalar component
$S \equiv {\rm e}^{-2\phi}  + ia$\ of the
chiral superfield ${\cal S}$   with   K\" ahler potential
\begin{equation}
K({\cal S},\overline {\cal S}) = -\alpha\,{\rm ln}({\cal S} + \overline
{\cal S})\ .\label{kahl}
\end{equation}
and  no superpotential ($W({\cal S}) = 0$).  In addition, ${\cal S}$ couples
 with a linear coupling to  the  kinetic energy of the Yang-Mills  superfield
(see
the bosonic part of the Lagrangian below).
 Note, that in $N =1$ supergravity theory  the value of  $\alpha$ in
(\ref{kahl}) is
an arbitrary positive constant.\footnote{ The Lagrangian with
the chiral superfield  ${\cal S}$  is  at the classical level equivalent  to
the
one with the linear supermultiplet ${\cal L}$, where  ${\cal S}$
 and  ${\cal L}$ are related through the duality transformation (see, {\it
e.g.},  Ref. \cite{ ABGG} and references therein for details). In the following
we
also assume that the particular $N=1$  Lagranigian is free from the mixed K\"
ahler-Lorentz anomalies (see, {\it e.g.}, Ref. \cite{CO} and references therein
for
details).}
For special values of  $\alpha$  one recovers the examples of
$N=1$ supergravity models with the corresponding wall solutions  mentioned
above. $\alpha=0$ corresponds   to the ``ordinary'' $N=1$ supergravity theory
(without the dilaton)  with the  extreme ordinary domain wall solutions.
$\alpha=1$
corresponds to the  effective $N=1$ supergravity theory of the tree level
$4d$ superstring vacua with extreme stringy domain wall solutions.
$\alpha=n\  (n=2,3,...)$ corresponds the case of
$N=1$ supergravity
theory  described with $n$ ``stringy'' dilatons and the extreme
wall solutions with the  naked singularity.
In the following we shall, however,  keep the value of  $\alpha$ arbitrary and
examine the nature of domain wall solution
 for the whole class of such supergravity models.

Along with   ${\cal S}$,  we introduce a
 chiral superfield ${\cal T}$  with  K\" ahler potential  $\break
K_M({\cal T},\overline
{\cal T})$ and nonzero superpotential $W_M({\cal T})$ and whose
 scalar component  $T$  forms the wall.
The effective tree level action  can  be written in terms
of a separable K\" ahler potential
$K = K_M ({\cal T}, \overline{\cal T}) + K({\cal S} , \overline{\cal S} )$ and
superpotential $W=W_M({\cal T})$.
The scalar part of the tree level
Lagrangian is then  of the form:
\begin{eqnarray}
&{\cal {L}} =\sqrt{-g} [- {1\over 2} R + K_{T \overline{T}}
\partial _\nu T \partial^\nu \overline{T} +
\alpha \partial_\nu \phi \partial^\nu \phi
 - 2^{-\alpha}{\rm e}^{2\alpha\phi}\tilde V
-{1\over 2}{\rm e}^{-2\phi}F_{\mu\nu}F^{\mu\nu} ]\label{lag}
\end{eqnarray}
where
\begin{equation}
\tilde V = {\rm e}^{K_M} [K^{T \overline{T}} |D_T W_M|^2 -
(3-\alpha)|W_M|^2 ]\label{pot}
\end{equation}
is the part of the potential, which  depends only on the matter field $T$.
Here $K^{T \overline{T}} \equiv (K_{T \overline
{T}})^{-1} \equiv (\partial_T \partial_{\overline {T}}K)^{-1}$ and $D_T
W_M \equiv {\rm e}^{-K_M}\partial_{T}({\rm e}^{K_M} W_M)$.
We use the metric convention $(+---)$ and  set the   gravitational  constant
 $\kappa\equiv 8\pi G=1$. In Eq.(\ref{lag})
 we have already set the axion field $a = 0$ (which turns out
 to correspond to the solution of equation of motion). For the wall
solutions   we also turn off the gauge
fields, {\it i.e.}, $F_{\mu\nu}=0$.

 We are looking for domain walls between isolated supersymmetric minima of
$\tilde V$.
Note, that $\tilde V$ is
modified due to the presence
of  dilaton field $\phi$, which yields
an additional contribution ${\rm e}^K K^{S,\overline {S}} |D_{S}W|^2
= 2^{-\alpha} {\rm e}^{2\phi}{\rm e}^{K_M}|W_M|^2$ to $\tilde{V}$.
 For  supersymmetric minima, $D_{T} W_{M} = 0$ and
$\tilde{V} = (\alpha - 3){\rm e}^{K_M} |W_M|^2$.  Therefore, at such  minima
dilaton  $\phi$  screens the matter potential by
$2^{-\alpha}{\rm e}^{2\alpha\phi}$   as well as changes
an overall
scale factor of  the matter potential from $-3$ (for the ordinary
supersymmetric
vacuum) to $(-3+\alpha)$, thus  rendering the matter potential (\ref{pot})
less negative.

We shall concentrate on the  extreme  Type I  solutions\footnote{Type II and
type
III walls correspond to the walls with
 $W_M\ne 0$ on both sides of the wall.
 Type II walls  have $W_{M}$ traversing zero,
 and Type III walls have  $W_{M} \ne 0$
everywhere. See Refs. \cite{CGR}, \cite{C}, and \cite{CY} for further details.}
  corresponding to the static, planar (say, in the $(x,y)$ plane located at
$z\sim 0$) where on one side (say, $z>0$) of the wall
the supersymmetric vacuum  corresponds to  $W_M(T)=0$ and
on the other side ($z<0$)of the wall the supersymmetric vacuum has  $W_M(T)\ne
0$.
We start with the metric
{\it Ansatz} for planar, static domain wall
solutions:
\begin{equation}
ds^2 = A(z)(dt^2 - dz^2 - dx^2 - dy^2),\label{met}
\end{equation}
and  the scalar fields $T(z)$, and $\phi(z)$ depend only on $z$.  Using a
technique of the generalized Israel-Nester-Witten  form developed in
Ref. \cite{CGR} for the
study of supergravity walls, one obtains the following  Bogomol'nyi bound
for the
energy density $\sigma$  of the planar domain wall configuration:
\begin{eqnarray}
&\sigma - |C| =
\int^{\infty}_{-\infty} [ -\delta _{\varepsilon}
\psi^{+} _{i}g^{ij} \delta _{\varepsilon} \psi _{j} + K_{T \overline {T}}
\delta _{\varepsilon}\chi ^+ \delta _{\varepsilon} \chi +
 K_{S
\overline{S}} \delta _{\varepsilon}\eta^+ \delta _{\varepsilon} \eta
]dz \geq 0. \label{bog}
\end{eqnarray}
This bound is saturated iff the supersymmetry variations $\delta _{\varepsilon}
\psi _{\mu}$, $\delta _{\varepsilon}\chi$, and $\delta _{\varepsilon} \eta $
of the fermionic partners of the fields $g_{\mu \nu}$, $T$ and $S$,
respectively, vanish.  For this case, one has {\it supersymmetric} bosonic
backgrounds, and the metric and scalar fields satisfy coupled first order
differential equations (self-dual or Bogomol'nyi equations):\footnote{The above
equations can be viewed as ``square roots'' of the corresponding Einstein and
Euler-Lagrange equations; they provide a particular solution of the equations
of motion which saturate the Bogomol'nyi bound (\ref{bog}) . The
existence  of such  static wall solutions is due to the constrained form of the
matter potential in $N=1$ supergravity theory. Note also, that in the thin
wall approximation one can explicitly solve the  Einstein
equations for $A(z)$  and  the Euler-Lagrange equation for
$\phi(z)$  outside the wall and then match the solution
 for $A(z)$ and $\phi(z)$ across the wall region. See Ref. \cite{CY}
for an explicit form of such equations. }
\begin{eqnarray}
&0={\rm Im} (\partial _{z}T {{D_{T}W_{M}} \over {W_M}})\nonumber \\
& \partial_{z} T = - (2^{-\alpha}A{\rm e}^{2\alpha\phi})^{1/2}
{\rm e}^{K_M /2} |W_M| K^{T \overline{T}} _M
{{D_{\overline{T}}\overline{W} _M}
\over {\overline{W} _M}} \nonumber \\
&\partial_{z} {\rm ln} A = 2 (2^{-\alpha}A{\rm e}^{2\alpha \phi})^{1/2}
{\rm e}^{K_{M}/2}|W_{M}|\nonumber \\
&\partial _{z}\phi = - (2^{-\alpha}A{\rm e}^{2\alpha \phi})^{1/2}
{\rm e}^{K_{M}/2}|W_{M}| \  .\label{boge}
\end{eqnarray}
The topological charge $|C|$, which determines the energy density of the
domain wall   can be determined explicitly in the thin wall
approximation.  Then in the wall region ($z \sim 0$) the matter
field $T$ is a quickly varying function, resembling   a step function centered
 at the wall,  while the metric $A(z)$ and
$\phi (z)$ fields  vary slowly.  With the choice  $A(0) = 1$ and the
boundary conditions $\phi(0) = \phi _{0}$ ,  one
obtains \cite{ CGR},\cite{C},\cite{CY}
\begin{equation}
\sigma= |C| \equiv 2 |{\rm e}^{K\over 2}W|_{z=0^-}= 2^{1-{\alpha \over 2}}{\rm
e}^{\alpha
\phi _{0}}|{\rm e}^{{K_M} \over 2}
W_M|_{z=0^-}\label{sig}
\end{equation}
  Here the subscript $0^-$ refers to the side of the wall
with $W_M(T)\ne 0$.

The first two equations in (\ref{boge}) describe the evolution of the
matter field $T = T(z)$ with $z$.  The first equation  is the ``geodesic''
equation \cite{CGR} for the complex $T$ field  and can be
always satisfied for Type I walls.  It is the same as for all
supergravity walls.  The third and fourth equations in
(\ref{boge}) for the conformal factor $A(z)$ and the real scalar
fields $\phi(z)$ imply:
\begin{equation}
A_s(z)\equiv A(z){\rm e}^{2\phi(z)} ={\rm e}^{2\phi_0}\ .\ \label{ap}
\end{equation}
Note, that this equation is true {\it everywhere} in the domain wall
background.  It implies that there is the choice (\ref{ap}) of a frame
where  the  metric is {\it flat}.

Here we concentrate on the explicit form of
solutions in the thin wall appro-ximation:\footnote{Numerical solutions of Eqs.
(\ref{boge})
for  a wall of any thickness  have  the same qualitative features.}
\begin{eqnarray}
&A(z) = 1,\ \ \ \ \ \ \ \ \ \ \
\ \ \ \ \ \ \ \ \ \ \ \ \ \ \ \ \ \ \ \ \ \ \ \ \ \ \ \ \ \
  z >0\ ;\nonumber \\
&A(z)
= \left[1 - {1\over 2}(\alpha - 1)|C||z|\right]^{2 \over {\alpha-1}}\ ,
 \ \alpha\ne
1\  ,   \ z<0\ ;\nonumber \\
 &A(z)={\rm e}^{-|C||z|}\ ,\ \ \ \ \ \ \ \ \
\ \ \ \ \ \ \ \ \ \ \ \ \ \ \alpha=1 \ , \ z<0\ ,
 \label{mets}
\end{eqnarray}
and the dilaton field satisfying Eq. (\ref{ap}) .
Thus, on one side of the wall ($z>0$)
the space-time is Minkowski; both the conformal factor $A(z)$ and  the dilaton
$\phi$ are constant.  However, on the other side ($z<0$)  $A(z)$
and $\phi(z)$ change its value. There the curvature is the form:
\begin{eqnarray}
 &R = {3\over 2} (2 - \alpha)
|C|^2\left[1 - {1\over 2}(\alpha - 1)|C||z|\right]^{- {{2\alpha}
\over{\alpha-1}}}\
,\ \alpha\ne 1\ ;\nonumber \\
&R = {3\over 2}|C|^2{\rm e}^{-|C||z|}\ ,\ \ \ \ \ \ \ \ \ \ \ \ \ \
\ \ \ \ \ \ \ \ \ \ \ \ \ \ \ \ \ \ \ \ \ \ \alpha=1\ .
\label{cur}
\end{eqnarray}
Such walls therefore act as ``windows'' from the
Minkowski space-time into the new type of supersymmetric space-times
whose  nature depends crucially on the value of the parameter $\alpha$:

\begin{itemize}
\item{$\alpha=0$, corresponds to the case of ordinary  supergravity
 walls, {\it i.e.},   the dilaton field is absent.  The induced
space-time  on $z<0$ side is    anti-de Sitter with the metric (\ref{met})
 in horospherical coordinates.  At $(t=\infty ,z=-\infty,)$ there is a
Cauchy horizon  with the zero surface gravity (${\cal K}\equiv
{1\over 2}\partial_z [{\rm
ln}A(z)]_{z=-\infty}=0 $)   and thus zero temperature ($T\equiv{\cal
K}/(2\pi)=0$) \cite{CDGS} . The geodesic
 extensions of the space-time was given  in
 Refs. \cite{CDGS} and \cite{GIBBIII} .
The most symmetric geodesic extension (see Figure 1) comprises of
a  system of  an infinite lattice
of semi-infinite Minkowsi space-times  separated by an anti-de Sitter core. }
 \begin{figure}[p]
\iffiginclude
\psfig{figure=marseille21.eps,width=84mm}
\fi
\caption{Penrose-Carter diagram in the $(z,t)$ direction  for the most
symmetric
geodesic  extension of
the extreme Type I ordinary domain wall ($\alpha=0$) comprises of
a  system of  an infinite lattice
of semi-infinite Minkowsi space-times ($M$) separated by an anti-de Sitter
core.
The  compact null coordinates
define the axes: $u,v = 2\tan^{-1}(t \mp z)$.
These coordinates can be smoothly extended across the
nulls separating the diamonds. The domain  wall region
is denoted with the thin lines.
Cauchy horizons (dashed lines) are the nulls separating
the anti-de Sitter  patches.}
\end{figure}
\vskip3mm
\item{
$\alpha< 1$ corresponds to the walls where curvature (\ref{cur})
blows up at $z=-\infty $. In a frame defined by (\ref{ap}) , the metric
is flat everywhere; in  this frame
the $z=-\infty$ point corresponds to the event  horizon. Thus,
the  singularity is covered by the horizon.  The associated temperature $T=0$.
The Carter-Penrose diagram
in the $(z,t)$ direction  for such walls
is given in  Figure 2.}
\begin{figure}[p]
\iffiginclude
\psfig{figure=marseille23.eps,height=60mm}
\fi
\caption{ Penrose-Carter diagram in the $(z,t)$ plane for
 extreme  Type I dilatonic domain wall  with $0<\alpha\le 1$ corresponds to the
the wall (denoted by a thin line) separating the semi-infinite Minkowski
space-time ($M$) and the space-time
with the varying dilaton field and  the null singularity
covered by the horizon (jagged line).
The compactified null coordinates
are defined in Figure 1. }
\end{figure}
\item{
$\alpha=1$ corresponds to the case  of stringy dilatonic wall
of $4d$ tree level string vacua.
The wall interpolates between the constant dilaton vacuum
(Minkowski-space-time) and   the linear dilaton vacuum, {\it i.e.},
$\phi=|C||z|/2$.  Note, that now the frame (\ref{ap}) with the flat metric
is the sigma model frame  (string frame) of the string theory.
In other words, string does not see the wall.
 At   $z=-\infty$   the string frame
metric has a horizon, while  in  the Einstein frame the curvature (\ref{cur})
 (and the dilaton) blow up \cite{C} .  Again, the
singularity is covered by the horizon
 and the  global space-time is
 the same as the one for the walls with $\alpha< 1$
(see Figure 2).  However, the temperature
associated with the horizon is now finite, {\it i.e.},
 $T=|C|/(4\pi)$.}
\item{
 $\alpha>1$ corresponds to the case where  the metric becomes singular
at a {\it  finite} coordinate
distance $|z|_{sing} = {2|C|/ (\alpha - 1)}$ from the wall.
  For $\alpha \ne 2$, the curvature  (\ref{cur})
blows-up  at $|z|_{sing}$. For $\alpha =2$, $R = 0$ but $R_{\mu \nu}R^{\mu \nu}
= \infty$ at the singularity.   Clearly, even in the frame (\ref{ap}) with the
 flat metric
$|z|_{sing}$  is a finite affine parameter way. Thus,
the singularity is naked.  The Carter-Penrose diagram
in the $(z,t)$ plane for this type of dilatonic
 domain walls  is given in  Figure 3.  In addition, the surface gravity ${\cal
K}$ blows-up at this point, and thus $T=\infty$.}
\end{itemize}
\vskip8mm
\begin{figure}[p]
\iffiginclude
\hbox{\hfill\psfig{figure=marseille22.eps,height=60mm}}
\fi
\caption{ Penrose-Carter diagram in the $(z,t)$ plane for
 extreme  Type I dilatonic domain wall  with $\alpha>1 $ corresponds to the
the wall (denoted by a thin line) separating the semi-infinite Minkowski
space-time ($M$) and the space-time with the varying dilaton field and the
naked (planar) singularity (jagged line). The
compactified null coordinates are $u,v = 2\tan^{-1}[t \mp (z+|z|_{sing})]$.}
\end{figure}
\vskip8mm
Extreme stringy
dilatonic walls ($\alpha=1$, $T=|C|/(4\pi)$)   therefore serve as a dividing
line between extreme walls ($\alpha<1$, $T=0$)
with  the (planar) singularity covered by the horizon  and the extreme
walls ($\alpha>1$, $T=\infty$) with the naked  (planar) singularity.
An important outstanding  problem is to find the  corresponding
non-extreme and ultra-extreme wall solutions.
In addition, quantum corrections  and stability of such configurations has to
be addressed.\footnote{It can be shown, by applying Killing spinor
identities \cite{KO}
that the extreme wall solutions do not acquire  quantum corrections,
which respect $N=1$ supersymmetry.}

The global  space-time of the above walls bears  striking similarities
to the one  of the corresponding extreme magnetically charged black
holes with a general dilaton coupling \cite{GM},\cite{HW} .
They are spherically symmetric solutions of the
Lagrangian (\ref{lag})  without matter fields $T$ ($V\equiv
0$), however, with non-zero gauge fields $F_{\mu\nu}\ne 0$.  In this case
 a  standard form
of the Lagrangian (\ref{lag}) is obtained by  redefining
 the  dilaton  field $\tilde \phi\equiv\sqrt{\alpha}\phi$; then the
 kinetic energy of
$\tilde \phi$ is normalized  and the corresponding coupling  gauge
 field kinetic energy is
of the form $-{1\over 2}{\em e}^{-2g\tilde \phi}F_{\mu\nu}F^{\mu\nu}$, where
$g=1/\sqrt{\alpha}$.\footnote{Note that
Lagrangian (\ref{lag}) is a
bosonic part of the $4d$  $N=1$ supergravity Lagrangian.  For
magnetically charged dilatonic black holes only
the cases with $g=1$ (arising from  the string
theory) and
$g= \sqrt 3$ (arising \cite{GA} from the $5d$ Kaluza-Klein theory)
 correspond to solutions of $N=4$ and $N=8$ supergravity Lagrangian,
respectively. See Refs.
\cite{KLOPP} and \cite{K} for a detailed discussion of  $N=4$
supersymmetry for a class of dilatonic charged
black holes.}  Extreme solutions with
$g=0$  correspond to the extreme  Reissner-Nordstr\" om black holes with
the same  global space-time  structure in the $(r,t)$ direction  as
the Type I supergravity  walls  in the $(z,t)$ direction
(see Figure 1). In the latter case, however,
the  time-like singularity is replaced by the  wall. In a very analogous
manner as in the case of extreme walls,  extreme charged
stringy dilatonic black holes ($g=1$,  $T=M/(8\pi)$)  serve as a
 dividing line \cite{GM} between
extreme charged dilatonic black holes ($g<1$, $T=0$) with the
singularity covered by the horizon and those ($g>1$, $T=\infty$)  with the
naked singularity.
\vskip4mm
{\bf\noindent Acknowledgments}
The presented  work   has been done in part in  collaboration with D.
Youm. The work is supported by  U.S. DOE  Grant No. DOE-EY-76-02-3071.


\begin{thebibliography}{99}
\bibitem[1]{CGR} {M. Cveti\v c, S. Griffies and S.-J. Rey, Nucl. Phys. {\bf
B381},
301 (1992).}
\bibitem[2]{CG}{M. Cveti\v c and S. Griffies, Phys. Lett. {\bf B285}, 27
(1992).}
\bibitem[3]{CDGS}{M. Cveti\v c, R. L. Davis, S. Griffies and H. H. Soleng,
 Phys. Rev. Lett. {\bf 70}, 1191 (1993).}
\bibitem[4]{GIBBIII}{G. Gibbons, Nucl. Phys. {\bf 394}, 3 (1993).}
\bibitem[5]{CGS}{M. Cveti\v c, S. Griffies and H. Soleng, Phys. Rev. Lett. {\bf
71},
670 (1993).}
\bibitem[6]{CGSII}{M. Cveti\v c, S. Griffies and
H. Soleng, Phys. Rev. {\bf48D }, 2613 (1993).}
\bibitem[7]{C}{M. Cveti\v c, Phys. Rev. Lett. {\bf 71}, 815 (1993).}
\bibitem[8]{GA}{G. Gibbons, Nucl. Phys. {\bf B207}, 337 (1982).}
\bibitem[9]{GM}{G. Gibbons and K.
Maeda, Nucl. Phys. {\bf B298}, 741 (1988).}
\bibitem[10]{GHS}{D. Garfinkle, G. Horowitz and A.
Strominger, Phys. Rev. {\bf D43}, 3140 (1991), {\bf D45},
3888 (1992){\bf E}.}
\bibitem[11]{H}{ G. Horowitz,
{\it The Dark Side of the String Theory: Black Holes and
Black Strings}, Santa Barbara preprint, UCSBTH-92-32 (October 1992),
hep-th/9210119.}
\bibitem[12]{CY}{M. Cveti\v c and D. Youm,
{\it Cosmic Censorship Violation for a Class
of Supersymmetric Solitons}, U. of Pennsylvania preprint, UPR-593-T (November
1993), hep-th/9311176. }
\bibitem[13]{ABGG}{ P. Adamietz, P. Binetruy,  G.Girardi and R. Grimm, Nucl.
Phys.
{\bf B401}, 257 (1993).}
\bibitem[14]{CO}{ G. Lopes Cardoso and B.
Ovrut, {\it Supersymmetric Calculation of Mixed K\" ahler-Gauge  and Mixed
K\" ahler-Lorentz Anomalies}, CERN preprint
CERN-TH.6961/93 (August 1993), hep-th/9308066.}
\bibitem[15]{KO}{R. Kallosh
and T. Ort\' in, {\it Killing Spinor Identities}, Stanford  preprint
SU-ITP-93-16 (June 1993), hep-th/9306085.}
\bibitem[16]{HW}{ C. Holzhey  and F. Wilczek, Nucl. Phys. {\bf B380}, 447
(1992).}
\bibitem[17]{KLOPP}{R. Kallosh, A. Linde, T. Ort\' in, A. Peet and A. Van
Proeyen,
Phys. Rev. {\bf D 46}, 5278 (1992).}
\bibitem[18]{K}{R. Kallosh,{\it Supersymmetry and Black Holes},
Stanford   preprint
PRINT-93-0504 (March 1993), hep-th/9306095.}
\end{thebibliography}
\end{document}